\documentclass[preprint,seceq]{jpsj2}
\usepackage{graphicx}

\def\runtitle{Many-Body Effects on Transmission through a Tunnel Junction}
\def\runauthor{Toshihiro \textsc{Kubo} and Arisato \textsc{Kawabata}}

\title{%
Many-Body Effects on the Transmission Probability\\
through a Tunnel Junction in a Strong Magnetic Field
}

\author{%
Toshihiro \textsc{Kubo}${^1,}$\thanks{E-mail: j1202701@ed.kagu.tus.ac.jp} and Arisato \textsc{Kawabata}$^{2,}$\thanks{E-mail: arisato.kawabata@gakushuin.ac.jp}
}

\inst{%
$^1$Department of Physics, Faculty of Science, Tokyo University of Science,\\
1-3 Kagurazaka, Shinjuku-ku, Tokyo 162-8601\\
$^2$Department of Physics, Gakushuin University, 1-5-1 Mejiro, Toshima-ku, Tokyo 171-8588, Japan
}

\recdate{\today}

\abst{%
We investigate effects of electron-electron interaction on the transmission probability of electrons through a tunnel junction in a strong magnetic field. We start with the Hartree-Fock approximation, and  we show that the coulomb interaction, which gives rise to the divergence of Fock correction, should be replaced by the dynamically screened interaction. We also show that we should use the bare coulomb interaction for Hartree term. We take into account higher order contributions using a simple renormalization group approach. The temperature dependence of the transmission probability is qualitatively similar to that of a one-dimensional system.
}

\kword{%
Friedel oscillations, strong magnetic field, one-dimensional system, screened coulomb interaction, poor man's scaling
}

\begin{document}
\sloppy
\maketitle

\section{Introduction}

Many-body effects on the electron transport in one-dimensional systems have been studied by several authors \cite{rf:1,rf:2,rf:3,rf:4}. The transport of interacting spinless Fermions through a single barrier was studied by Kane and Fisher \cite{rf:1} and the case of spin-1/2 Fermions was investigated by Furusaki and Nagaosa \cite{rf:2}. They found that the transmission probability vanishes with a power-law as $T\to 0$. They treated the problem within the framework of the Tomonaga-Luttinger liquid theory \cite{rf:6,rf:7}, i.e., they neglected the backward scatterings in the electron-electron interaction.

After that, Matveev \textit{et al}. developed a different treatment in which the effects of the backward scattering can be incorporated \cite{rf:3,rf:4}. They started with the Hartree-Fock approximation and treated the logarithmic singularity by a simple renormalization group theory. They found that the temperature dependence of the transmission probability does not obey a simple power-law. According to them, Friedel oscillations of the electron density induced by the barrier potential give an essential effect on the electrical conduction in one-dimensional systems. 

Those theories give qualitatively the same results, as for the vanishing transmission probabilities at zero-temperature. Experimentally, this tendency was observed in quantum wires \cite{rf:9}. In this experiment, however, electrons are probably scattered by the disorder such as the fluctuation of the width of the wire, and the situation is not as simple as the case of a single barrier \cite{rf:10,rf:11}. In fact, in spite of the development of the technique, it is still hard to verify the theoretical predictions using artificial one-dimensional systems, because the localization effects also contribute to the decrease of the conductance at low temperatures.

\begin{figure}[htbp]
  \begin{center}
    \includegraphics[scale=0.5]{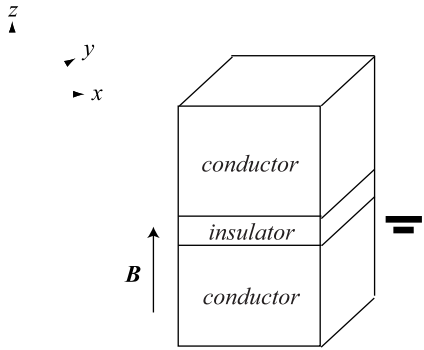}
  \end{center}
  \caption{Tunnel junctions in a strong magnetic field. Magnetic field  $\mib{B}=(0,0,B)$ is perpendicular to the insulator thin film. }
  \label{Fig1}
 \end{figure}

In this paper, we investigate the transport of electrons through a tunnel junction with a strong magnetic field perpendicular to it (see Fig.~\ref{Fig1}):  The system is one-dimensional like when only the lowest Landau levels is occupied, and we can expect a similar effect. In fact, 
recently, Biagini \textit{et al.} and Tsai \textit{et al.} treated this problem and have shown that the transmission probability behaves like that of one-dimensional systems\cite{rf:16,rf:17}. 

In this system, measurement of the transmission probability is much easier than in one-dimensional systems. Moreover, one may be able to extract the interaction effects from the data by making use of the magnetic field dependence of the transmission probability. 

In order to do so, it is important to estimate the parameters which determine the temperature dependence of the transmission probability in approximations as good as possible. The important parameter are $\tilde{V}(0)$, and $\tilde{V}(2k_F)$, where $\tilde{V}(q)$ is the Fourier transform of the interaction potential $V(x)$ and $k_F$ is the Fermi wave number. For coulomb interaction, $\tilde{V}(0)$ is divergent, and in Refs. \citen{rf:16} and \citen{rf:17} the authors used the static screened coulomb interaction for $\tilde{V}(0)$, and $\tilde{V}(2k_F)$. This replacement is not appropriate, however, as will be shown in the following parts of this paper. We will see that we should use dynamical screened interaction for $\tilde{V}(0)$ and bare coulomb interaction for $\tilde{V}(2k_F)$.

In section~\ref{HFCTP} we present the calculations of Hartree-Fock correction to the transmission amplitude, and in section~\ref{HOC} we will show the way to take into account the higher order terms using renormalization group method. 

\section{Hartree-Fock Correction to Transmission Probability}\label{HFCTP}

\subsection{Non-interacting electrons}
First we investigate the transmission of a noninteracting 3D electron through tunnel junctions under a strong magnetic field. We ignore the spin degrees of freedom, for we assume that the electrons are fully polarized. We choose the $z$-axis of the coordinate system along the magnetic field, and use Landau gauge $\mib{A}(\mib{x})=(0,Bx,0)$ for the vector potential. Then the single-electron Hamiltonian is of the form
\begin{equation}
H_0=\frac{1}{2m}(\mib{p}+e\mib{A}(\mib{x}))^2+U(z), \label{ham1}
\end{equation}
where $U(z)$ is the barrier potential of the insulator thin film, and we assume that the potential barrier is localized around $z=0$ , i.e., $U(z)=0$ for $|z|>a$.

We consider the case when the electrons occupies only the lowest Landau level. It is realized if the magnetic field is strong enough so that
\begin{equation}
B>\frac{\hbar}{e}(2\pi^4n_e^2)^{1/3}, \label{mag}
\end{equation}
where $n_e$ is the electron density. Then, we can label the eigenstates of the Hamiltonian $H_0$ by a two-dimensional vector $\mib{k}=(k_y,k_z)$, and the energy eigenvalues $\epsilon_{\mib{k}}^0$ and the wave functions $\varphi_{\mib{k}}^0(\mib{x})$ are of the forms
\begin{subequations}
\begin{align}
&\epsilon_{\mib{k}}^0=\frac{1}{2}\hbar\omega_c+\frac{\hbar^2{k_z}^2}{2m}\,, 
\label{energy}\\
&\varphi_{\mib{k}}^0(\mib{x})=\phi_{k_y}^0(x,y)u_{k_z}^0(z)\,,
\label{wf0}
\end{align}
\end{subequations}
where
\begin{subequations}\label{ukt}
\begin{equation}
\phi_{k_y}^0(x,y)=\frac{1}{\pi^{1/4}{\ell_B}^{1/2}}\exp\left[-\frac{(x+k_y{\ell_B}^2)^2}{2{\ell_B}^2} \right]e^{ik_yy},
\hspace{1.5cm}
\end{equation}
\begin{align}
u_{k_z}^0(z)=\left\{
  \begin{array}{cc}
   e^{ik_zz}+r_0e^{-ik_zz},& (z<-a),\\
   t_0e^{ik_zz},& (z>a),
  \end{array}
\right.\ (k_z>0),
\label{uk1}
\end{align}
\begin{align}
u_{k_z}^0(z)=\left\{
  \begin{array}{cc}
    t_0e^{ik_zz},& (z<-a),   \\
    e^{ik_zz}+r_0e^{-ik_zz}, & (z>a),
  \end{array}
\right.\ (k_z<0).
\label{uk2}
\end{align}
\end{subequations}
Here $\omega_c=eB/m$, $\ell_B=\sqrt{\hbar/eB}$, and $t_0$ and $r_0$ are the transmission and reflection amplitudes through the barrier, respectively. We assume an infinite system and neglect the effects of the boundaries.

\subsection{First Born approximation}

Next we calculate the correction to the transmission probability due to the electron-electron interaction. We calculate it to the lowest order in the interaction within the Hartree-Fock theory. Using the electron field operator $\psi(\mib{x})$, we can write the many-body Hamiltonian ${\cal H}$ as
\begin{equation}
{\cal H}={\cal H}_0+{\cal H}_1, \label{ham2}
\end{equation}
with
\begin{align}
{\cal H}_0&=\int d\mib{x}\psi^{\dagger}(\mib{x})\left[\frac{1}{2m}\left(\mib{p}+e\mib{A}(\mib{x}) \right)^2+U(z) \right]\psi(\mib{x}),\label{ham3}\nonumber\\
&
\end{align}
and
\begin{align}
{\cal H}_1&=\frac{1}{2}\iint d\mib{x}d\mib{x}'\psi^{\dagger}(\mib{x})\psi^{\dagger}(\mib{x}')V(\mib{x}-\mib{x}')\psi(\mib{x}')\psi(\mib{x}),\label{ham4}\nonumber\\
&
\end{align}
$V(\mib{x})$ being the coulomb interaction potential, and if we do not write explicitly the ranges of the space integrations they mean integrations over the whole space region. We calculate the correction to the wave functions using Green's function method \cite{rf:12}.

The single-electron wave functions $\varphi_{\mib{k}}(\mib{x},\tau)$ can be written in the form
\begin{equation}
\left.
  \begin{array}{cc}
     \varphi_{\mib{k}}(\mib{x},\tau)=i\int d\mib{y}G_R(\mib{x},\tau;\mib{y},\tau_0)\varphi_{\mib{k}}(\mib{y},\tau_0),  &  \tau>\tau_0  \\
  \end{array}
\right.\label{wf3}
\end{equation}
in terms of the retarded Green's function $G_R(\mib{x},\tau;\mib{y},\tau_0)$. Assuming that the interaction is switched on at time $\tau_0$, we put $\varphi_{\mib{k}}(\mib{x},\tau_0)=\varphi_{\mib{k}}^0(\mib{x})e^{-i\xi_{\mib{k}}^0\tau_0/\hbar}$, $\xi_{\mib{k}}^0$ being the single-electron energy measured from the Fermi level, and later we will let $\tau_0\to -\infty$. The retarded Green's function $G_R(\mib{x},\mib{y},\omega)$ can be obtained from the Matsubara Green's function ${\cal G}(\mib{x},\mib{y},\omega_n)$ in terms of the analytic continuation
\begin{equation}
G_R(\mib{x},\mib{y},\omega)={\cal G}(\mib{x},\mib{y},-i\omega+\eta)\label{green1}
\end{equation}
where $\omega_n=\pi k_BT(2n+1)/\hbar$, $n$ being an integer, and $\eta\to +0$.
\begin{figure}[htbp]
  \begin{center}
    \includegraphics[scale=0.5]{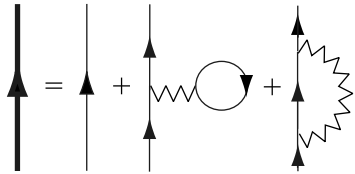}
  \end{center}
  \caption{Feynman diagram for the first Born approximation within Hartree-Fock theory: The thin solid lines indicate the Green's function without the interaction correction, and the wavy lines indicate the interaction.}
\label{Fig2}
\end{figure}
The unperturbed Matsubara Green's function ${\cal G}^0(\mib{x},\mib{y},\omega_n)$ is defined by
\begin{equation}
{\cal G}^0(\mib{x},\mib{y},\omega_n)=\int\frac{d\mib{k}}{(2\pi)^2}\frac{\varphi_{\mib{k}}^0(\mib{x}){\varphi_{\mib{k}}^0}^*(\mib{y})}{i\omega_n-\xi_{\mib{k}}^0/\hbar}.\label{green2}
\end{equation}
Hereafter, if we do not write explicitly the ranges of the wave number integrations, they mean integrations over the whole wave number region. The Feynman diagrams for ${\cal G}(\mib{x},\mib{y},\omega_n)$ to the lowest order in the interaction is shown in Fig.~\ref{Fig2}, and the expression corresponding to it is easily found to be
\begin{align}
{\cal G}(\mib{x},\mib{y},\omega_n)&={\cal G}^0(\mib{x},\mib{y},\omega_n)+\int d\mib{x}_1d\mib{x}_1'\nonumber\\
&\times{\cal G}^0(\mib{x},\mib{x}_1,\omega_n)\Sigma_{HF}(\mib{x}_1,\mib{x}_1'){\cal G}^0(\mib{x}_1',\mib{y},\omega_n),\label{green3}
\end{align}
where $\Sigma_{HF}(\mib{x}_1,\mib{x}_1')$ is the Hartree-Fock self-energy which is independent of the frequency $\omega_n$ and is given by
\begin{align}
\hbar\Sigma_{HF}(\mib{x}_1,\mib{x}_1')=&\delta(\mib{x}_1-\mib{x}_1')\int d\mib{x}_2V(\mib{x}_1-\mib{x}_2)\nonumber\\
&\times k_B T\sum_{n'}{\cal G}^0(\mib{x}_2,\mib{x}_2,\omega_{n'})\nonumber\\
&-V(\mib{x}_1-\mib{x}_1')k_B T\sum_{n'}{\cal G}^0(\mib{x}_1,\mib{x}_1',\omega_{n'}).\label{self1}
\end{align}

Then we obtain the following form for single-electron wave functions (see Appendix A)
\begin{align}
\varphi_{\mib{k}}(\mib{x})=&\varphi_{\mib{k}}^0(\mib{x})+\int d\mib{x}_1G_{\mib{k}}(\mib{x};\mib{x}_1)V_H(\mib{x}_1)\varphi_{\mib{k}}^0(\mib{x}_1)\nonumber\\
&-\iint d\mib{x}_1d\mib{x}_1'G_{\mib{k}}(\mib{x};\mib{x}_1)V_F(\mib{x}_1,\mib{x}_1')\varphi_{\mib{k}}^0(\mib{x}_1'),\label{wf4}
\end{align}
where 
\begin{equation}
G_{\mib{k}}(\mib{x};\mib{x}_1)\equiv\int\frac{d\mib{k}'}{(2\pi)^2}\frac{\varphi_{\mib{k}'}^0(\mib{x}){\varphi_{\mib{k}'}^0}^*(\mib
{x}_1)}{\xi_{\mib{k}}^0-\xi_{\mib{k}'}^0},\label{green4}
\end{equation}
\begin{equation}
V_H(\mib{x}_1)=\int d\mib{x}_2V(\mib{x}_1-\mib{x}_2)\rho(\mib{x}_2),\label{hpot1}
\end{equation}
and
\begin{equation}
V_F(\mib{x}_1,\mib{x}_1')=V(\mib{x}_1-\mib{x}_1')\int_{oc.}\frac{d\mib{k}}{(2\pi)^2}\varphi_{\mib{k}}^0(\mib{x}_1){\varphi_{\mib{k}}^0}^*(\mib{x}_1'),\label{fpot1}
\end{equation}
$\rho(\mib{x})$ being the electron density and the subscript $oc.$ means the integration over the occupied states, i.e., $|k_z|\le k_F$. As for the treatment of the singularity in eq. (\ref{green4}), the reader is referred to Appendix B.

To calculate the correction to the transmission probability we need only the asymptotic form of the $G_{\mib{k}}(\mib{x},\mib{x}_1)$ at $z\to +\infty$. From eq. (\ref{green4}), we obtain (see Appendix B)
\begin{align}
G_{\mib{k}}(\mib{x};\mib{x}_1)&=\frac{1}{2\pi i\hbar v_F{\ell_B}^2}\exp\left[-\frac{(x-x_1)^2+(y-y_1)^2}{4{\ell_B}^2}\right]\nonumber\\
&\times\exp\left[-\frac{i(x+x_1)(y-y_1)}{2{\ell_B}^2} \right]\nonumber\\
&\times\left\{
  \begin{array}{cc}
    t_0e^{ik_z(z-z_1)},& z_1<0,  \\
    e^{ik_z(z-z_1)}+r_0e^{ik_z(z+z_1)},& z_1>0,
  \end{array}
\right.
\label{green5}
\end{align}
with $v_F \equiv \hbar k_F/m$.
The electron density $\rho(\mib{x})$ in eq. (\ref{hpot1}) is calculated from the wave functions (\ref{wf0}) with (\ref{ukt}), and at large distances $|z|$ it behaves like
\begin{align}
\rho(\mib{x})&\equiv \int_{oc.}\frac{d\mib{k}}{(2\pi)^2}|\varphi_{\mib{k}}^0(\mib{x})|^2\nonumber\\
&\simeq \frac{1}{2\pi{\ell_B}^2}\cdot \left[\frac{|r_0|}{2\pi|z|}\sin(2k_F|z|+\arg r_0)+\frac{k_F}{\pi}\right]\nonumber\\
&\equiv \frac{1}{2\pi{\ell_B}^2}\cdot n(z).\label{density}
\end{align}
In the following  we will neglect the space independent part in $n(z)$ which gives a constant Hartree potential. Such oscillation of the electron density, i.e., the Friedel oscillation, gives an essential effect on the transmission probability, as in one-dimensional systems \cite{rf:3,rf:4}.

\subsection{Fock correction}\label{subsecFock}

First we calculate the Fock correction  $\varphi_{\mib{k}}^{(1F)}(\mib{x})$ to the wave function, i.e., the second term of the right side of eq.~(\ref{wf4})
\begin{align}
\varphi_{\mib{k}}^{(1F)}(\mib{x})=&-\iint d\mib{x}_1d\mib{x}_1'G_{\mib{k}}(\mib{x};\mib{x}_1)V_F(\mib{x}_1;\mib{x}_1')\varphi_{\mib{k}}^0(\mib{x}_1')\nonumber\\
=&-\iint d\mib{x}_1d\mib{x}_1G_{k_y}(x,y;x_1,y_1)G_{k_z}(z;z_1)\nonumber\\
&\times\left[V(\mib{x}_1-\mib{x}_1')\int_{oc.}\frac{d\mib{k}'}{(2\pi)^2}\varphi_{\mib{k}'}^0(\mib{x}_1){\varphi_{\mib{k}'}^0}^*(\mib{x}_1') \right]\nonumber\\
&\times\phi_{k_y}^0(x_1',y_1')u_{k_z}^0(z_1')\nonumber\\
=&-\iiiint dx_1dy_1dx_1'dy_1'G_{k_y}(x,y;x_1,y_1)\nonumber\\
&\times\left[\int\frac{dk_y'}{2\pi}\phi_{k_y'}^0(x_1,y_1){\phi_{k_y'}^0}^*(x_1',y_1') \right]\phi_{k_y}^0(x_1',y_1')\nonumber\\
&\times\iint dz_1dz_1'G_{k_z}(z;z_1)u_{k_z}^0(z_1')\nonumber\\
&\times\left[V(\mib{x}_1-\mib{x}_1')\int_{-k_F}^{k_F}\frac{dk_z'}{2\pi}u_{k_z'}^0(z_1){u_{k_z'}^0}^*(z_1') \right], \label{fwf1}
\end{align}
where $G_{k_y}(x,y;x_1,y_1)$ and $G_{k_z}(z;z_1)$ are defined in Appendix B.

As is shown in the Appendix C, the integration over $z_1$, $z_1'$, and $k_z'$ gives
\begin{align}
& \iint dz_1dz_1'G_{k_z}(z;z_1)\nonumber\\
&\times\left[V(\mib{x}_1-\mib{x}_1')\int_{-k_F}^{k_F}\frac{dk_z'}{2\pi}u_{k_z'}^0(z_1){u_{k_z'}^0}^*(z_1') \right]u_{k_z}^0(z_1')\nonumber\\
=&\frac{t_0|r_0|^2}{2\pi\hbar v_F}\tilde{V}(x_1-x_1',y_1-y_1';0)\ln\left(\frac{1}{|k_z-k_F|d} \right)\cdot e^{ik_zz},\label{fwf2}
\end{align}
where $\tilde{V}(x_1-x_1',y_1-y_1';q_z)$ is the Fourier transform of $V(\mib{x}_1-\mib{x}_1')$ with respect only to $z$ component of $\mib{x}_1-\mib{x}_1'$ and is defined by,
\begin{equation}
\tilde{V}(x_1-x_1',y_1-y_1';k_z)\equiv \int_{-\infty}^{\infty}dz_2'V(x_1-x_1',y_1-y_1',z_2')e^{-ik_zz_2'}.
\label{c6}
\end{equation}
and $d$ is the cut-off length. 

From Eqs. (\ref{fwf1}) and (\ref{fwf2}), we have
\begin{align}
\varphi_{\mib{k}}^{(1F)}(\mib{x})=&-\iiiint dx_1dy_1dx_1'dy_1'G_{k_y}(x,y;x_1,y_1)\nonumber\\
&\times\left[\int\frac{dk_y'}{2\pi}\phi_{k_y'}^0(x_1,y_1){\phi_{k_y'}^0}^*(x_1',y_1') \right]\phi_{k_y}^0(x_1',y_1')\nonumber\\
&\times\frac{t_0|r_0|^2}{2\pi\hbar v_F}\tilde{V}_s(x_1-x_1',y_1-y_1';0)\nonumber\\
&\times\ln\left(\frac{1}{|k_z-k_F|d} \right)\cdot e^{ik_zz}.\label{fwf3}
\end{align}
Thus the contribution from the Fock term $\varphi_{\mib{k}}^{(1F)}(\mib{x})$ is written as (see Appendix C)
\begin{align}
\varphi_{\mib{k}}^{(1F)}(\mib{x})&=-\alpha_2(B)t_0(1-|t_0|^2)\ln\left(\frac{1}{|k_z-k_F|d} \right)\nonumber\\
&\times\phi_{k_y}^0(x,y)e^{ik_zz},\label{fwf4}
\end{align}
where
\begin{equation}
\alpha_2(B) \equiv \frac{1}{2\pi^2\hbar v_F}\int_{0}^{\infty}q_{\perp}dq_{\perp}e^{-q_{\perp}^2{\ell_B}^2/2}\,\hat{V}(\mib{q}_{\perp})\,,
\label{alpha2}
\end{equation}
$\hat{V}(\mib{q}_{\perp})$ being the Fourier transform of interaction $V(\mib{x})$ with $\mib{q}_{\perp} \equiv (q_x, q_y, 0)$, and we obtain the following 1st-order correction to the transmission amplitude by the Fock term
\begin{equation}
t^{(1F)}=-\alpha_2(B)t_0(1-|t_0|^2)\ln\left(\frac{1}{|k_z-k_F|d} \right).\label{famp}
\end{equation}

It is to be noted that if $V(\mib{x})$ is coulomb interaction we have
\begin{equation}
\hat{V}(\mib{q})=\frac{e^2}{\epsilon q^2}\,,
\end{equation}
$\epsilon$ being the dielectric permittivity of the matter, and that the right hand side of eq.~(\ref{alpha2}) is divergent.
In this context, one of the authors (A.K.) reinvestigated the problem from a point of view different from those of the theories mentioned above \cite{rf:5}: Using the parquet diagram method developed by Bychkov \textit{et al} \cite{rf:8}, he has shown that, to be consistent, this singularity should be renormalized, too, and that the interaction potential should be replaced by the screened interaction within Random Phase Approximation (RPA) \cite{rf:13}.

As the screened interaction is frequency dependent, the second term of the right hand side of eq. (\ref{self1}) should be replaced by
\begin{equation}
-\sum_{n'}V(\mib{x}_1-\mib{x}_1', i\omega_n- i\omega_{n'}){\cal G}^0(x_1',x_1;\omega_{n'}) \,.\label{eqx}
\end{equation}
According to the arguments in Ref. \citen{rf:5}, relevant energy is $\hbar v_F q_z \approx \hbar\omega_n \approx k_B T$ and we have to replace $\hat{V}(\mib{q}_{\perp})$ in eq.~(\ref{alpha2}) with 
\begin{equation}
\lim_{q_z \to 0}\hat{V}_s(\mib{q}, \omega_n = v_F q_z)\,.
\label{limVqom}
\end{equation}
where frequency dependent screened interaction is given by eq.~(\ref{epsqom2}). Then we easily find that
\begin{equation}
\alpha_2(B)\equiv 2(\kappa\ell_B)^2\int_{0}^{\infty}dq
\frac{q e^{-q^2{\ell_B}^2/2}}{q^2 + \kappa^2 e^{-q^2{\ell_B}^2/2}}\,,
\label{fpara}
\end{equation}
$\kappa$ being given by eq.~(\ref{kappa}). We easily find that this $\kappa$ is smaller than that in Ref.~\citen{rf:17} by a factor $1/\sqrt{2}$.  

\subsection{Hartree correction}

Next we calculate the correction to the transmission amplitude by the Hartree term. We can write the contribution of the Hartree term $\varphi_{\mib{k}}^{(1H)}(\mib{x})$ as
\begin{align}
\varphi_{\mib{k}}^{(1H)}(\mib{x})=&\int d\mib{x}_1G_{\mib{k}}(\mib{x};\mib{x}_1)V_H(\mib{x}_1)\varphi_{\mib{k}}^0(\mib{x}_1)\nonumber\\
=&\int d\mib{x}_1G_{k_y}(x,y;x_1,y_1)G_{k_z}(z;z_1)\phi_{k_y}^0(x_1,y_1)\nonumber\\
&\times\left[\int d\mib{x}_2V(\mib{x}_1-\mib{x}_2)\rho(\mib{x}_2) \right]u_{k_z}^0(z_1)\nonumber\\
=&\iint dx_1dy_1G_{k_y}(x,y;x_1,y_1)\phi_{k_y}^0(x_1,y_1)\nonumber\\
&\times\iint \frac{dx_2dy_2}{2\pi{\ell_B}^2}
\int dz_1G_{k_z}(z;z_1)u_{k_z}^0(z_1)\nonumber\\
&\times\left[\int dz_2V(\mib{x}_1-\mib{x}_2)n(z_2) \right].
\label{hwf1}
\end{align}
Here we should not replace $V(\mib{x})$ with the screened interaction according to the following reason.

\begin{figure}[htbp]
  \begin{center}
    \includegraphics[scale=0.5]{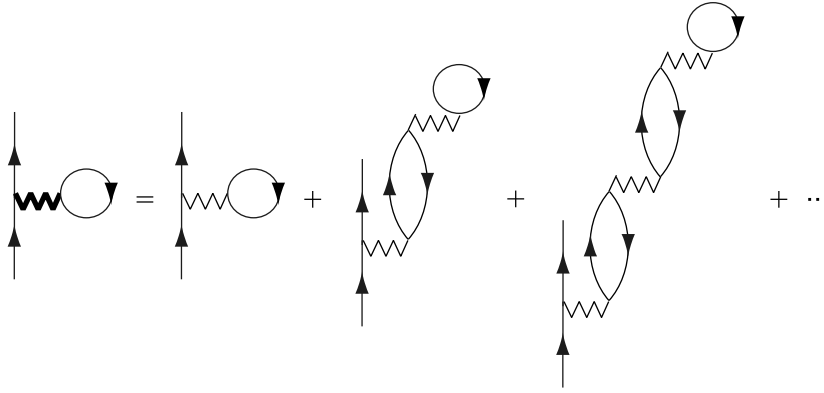}
  \end{center}
  \caption{The Hartree correction in terms of the screened coulomb interaction}
\label{f10}
\end{figure}

\begin{figure}[htbp]
  \begin{center}
    \includegraphics[scale=0.5]{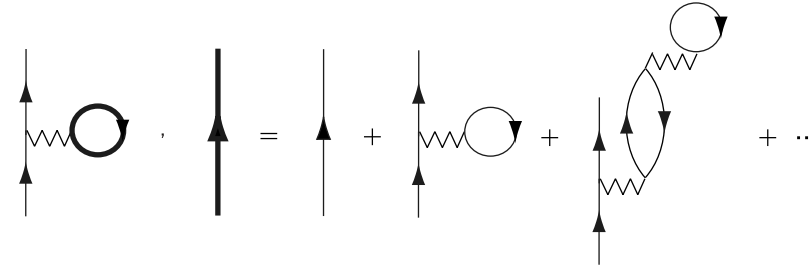}
  \end{center}
  \caption{The Hartree correction in terms of the one-electron Green's function}
  \label{f11}
\end{figure}

The above mentioned replacement can be expressed by Feynman diagrams in Fig. \ref{f10} as the renormalization of the interaction. On the other hand, this term can be interpreted as the correction to the one-electron Green's function as is shown in Fig. \ref{f11}.
In the section \ref{HOC} we take into account the higher order correction, which can be interpreted as the corrections to the one-electron states. Therefore, if we replace the coulomb interaction with the screened interaction, it gives rise to the overcounting of the corrections.

In other words, the effects of the Hartree terms are essentially the screening of the barrier potential, and the screening should not be counted twice. Thus, in the Hartree term, we should keep the bare coulomb potential in eq. (\ref{self1}).

As is shown in the Appendix C, the integration over $z_1$ and $z_2$ in eq.~(\ref{hwf1}) gives
\begin{align}
& \int dz_1G_{k_z}(z;z_1)u_{k_z}^0(z_1)\nonumber\\
&\times\left[\int dz_2V(\mib{x}_1-\mib{x}_2)n(z_2) \right]\nonumber\\
=&\frac{t_0|r_0|^2}{2\pi\hbar v_F}\tilde{V}(x_1-x_2,y_1-y_2;2k_F)\nonumber\\
&\times\ln\left(\frac{1}{|k_z-k_F|d} \right)\cdot e^{ik_zz}.\label{hwf2}
\end{align} 
From Eqs. (\ref{hwf1}) and (\ref{hwf2}), we have
\begin{align}
\varphi_{\mib{k}}^{(1H)}(\mib{x})=&\iint dx_1dy_1G_{k_y}(x,y;x_1,y_1)\nonumber\\
&\times\left[\iint dx_2dy_2\frac{1}{2\pi{\ell_B}^2} \right]\phi_{k_y}^0(x_1,y_1)\nonumber\\
&\times\frac{t_0|r_0|^2}{2\pi\hbar v_F}\tilde{V}(x_1-x_2,y_1-y_2;2k_F)\nonumber\\
&\times\ln\left(\frac{1}{|k_z-k_F|d} \right)\cdot e^{ik_zz},\label{hwf3}
\end{align}
hence the contribution from the Hartree term $\varphi_{\mib{k}}^{(1H)}(\mib{x})$ is written as (see Appendix C)
\begin{align}
\varphi_{\mib{k}}^{(1H)}(\mib{x})&=\alpha_1(B)t_0(1-|t_0|^2)\ln\left(\frac{1}{|k_z-k_F|d} \right)\nonumber\\
&\times\phi_{k_y}^0(x,y)e^{ik_zz},\label{hwf4}
\end{align}
where
\begin{equation}
\alpha_1(B)\equiv \frac{\kappa^2}{4{k_F}^2}.\label{hpara}
\end{equation}
Thus we obtain the following 1st-order correction to the transmission amplitude by the Hartree term.
\begin{equation}
t^{(1H)}=\alpha_1(B)t_0(1-|t_0|^2)\ln\left(\frac{1}{|k_z-k_F|d} \right).\label{hamp}
\end{equation}

Eqs. (\ref{famp}) and (\ref{hamp}) give the 1st-order correction to the transmission amplitude within Hartree-Fock theory
\begin{equation}
t^{(1)}=-\alpha(B)t_0(1-|t_0|^2)\ln\left(\frac{1}{|k_z-k_F|d} \right),\label{hfamp}
\end{equation}
where dimensionless parameter $\alpha(B)$ of the electron-electron interaction is
\begin{align}
\alpha(B)&=\alpha_2(B)-\alpha_1(B),\label{hfpara}
\end{align}
and the 1st-order correction to the transmission probability is given by
\begin{equation}
{\cal T}^{(1)}=-2\alpha(B){\cal T}_0(1-{\cal T}_0)\ln\left(\frac{1}{|k_z-k_F|d} \right).\label{prob}
\end{equation}

Since this is the result obtained by perturbation theory, it is applicable as long as
\begin{equation}
\alpha(B)\ln\left(\frac{1}{|k_z-k_F|d} \right)\ll 1.\label{condition}
\end{equation}
However eq. (\ref{condition}) is no longer valid  at low temperature since only electrons of $k_z\simeq k_F$ contribute electric conduction at low temperatures. Thus we have to take into account the higher order contributions in the interaction.

\section{Higher order contributions}\label{HOC}

In order to include the higher order corrections, we use a simple renormalization group (RG) approach called the poor man's scaling developed by Anderson for the Kondo problem \cite{rf:14}.
\begin{figure}[htbp]
  \begin{center}
    \includegraphics[scale=0.4]{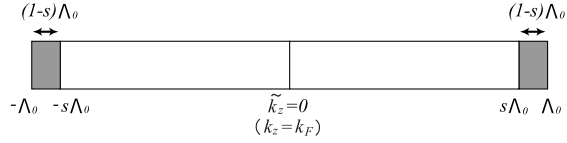}
  \end{center}
  \caption{The reduction of the cutoff $\Lambda_0$. $\tilde{k}_z$ is the wave number measured from $k_F$.}
\label{PoorRGT}
\end{figure}

We assume that only electrons in the strip of the wave number $k_z$ of halfwidth $\Lambda_0=1/d$ near the Fermi wave number $k_F$ contribute to the correction (\ref{prob}) (see Fig.~\ref{PoorRGT}). In this strip we linearize the dispersion relation of electron energy: 
\begin{equation}
\xi_{\mib{k}}^0=\hbar v_F(|k_z|-k_F).
\label{dispersion}
\end{equation}
Note that $d$ is the cutoff of integrations not only due to the validity of eq. (\ref{density}) but also due the validity of the linearized dispersion.

Next we reduce the cutoff $\Lambda_0$ to $s\Lambda_0$ with $1-s\ll 1$ (see Fig.~\ref{PoorRGT}). If we simultaneously renormalize the transmission probability ${\cal T}_0$ so that the effects of the states excluded by this RG transformation are taken into account, the new problem by this RG transformation is equivalent to the original problem. The change in ${\cal T}_0$ found in the first Born approximation is (see Appendix E)
\begin{equation}
\delta{\cal T}=-2\alpha(B){\cal T}_0(1-{\cal T}_0)\ln\left(\frac{E_0}{E} \right),\label{rgprob1}
\end{equation}
where $E\equiv\hbar v_Fs\Lambda_0,E_0\equiv\hbar v_F\Lambda_0$ are respectively the electron energy measured from Fermi level.

We apply the RG transformation again, reducing the bandwidth $E\to sE$ step by step. During each step of rescaling the cutoff, transmission probability ${\cal T}$ is renormalized according to eq. (\ref{rgprob1}) with ${\cal T}_0$ being substituted by the modified ${\cal T}$ from the previous step. The effect of these renormalizations may be found as a solution of the differential equation (RG equation)
\begin{equation}
\frac{d{\cal T}(E)}{d\ln(E_0/E)}=-2\alpha(B){\cal T}(E)(1-{\cal T}(E)).\label{rgeq}
\end{equation}

If we integrate this RG equation from $E=E_0$ to $E={\cal E}$ with the initial condition ${\cal T}|_{E=E_0}={\cal T}_0$, the transmission probability becomes
\begin{equation}
{\cal T}({\cal E})=\frac{{\cal T}_0({\cal E}/E_0)^{2\alpha(B)}}{{\cal R}_0+{\cal T}_0({\cal E}/E_0)^{2\alpha(B)}},\label{rgprob2}
\end{equation}
where ${\cal R}_0=1-{\cal T}_0$. At low temperature ${\cal E}$ in eq. (\ref{rgprob2}) should be replaced by $k_BT$ and the temperature dependence of the transmission probability is found to be
\begin{equation}
{\cal T}(T)=\frac{{\cal T}_0(k_BT/E_0)^{2\alpha(B)}}{{\cal R}_0+{\cal T}_0(k_BT/E_0)^{2\alpha(B)}}.\label{rgprob3}
\end{equation}
Our result is of the same form as that of one-dimensional electron systems except for the parameter of the electron-electron interaction.

\subsection{Temperature dependence of the transmission probability}

\begin{figure}[htbp]
  \begin{center}
    \includegraphics[scale=0.8]{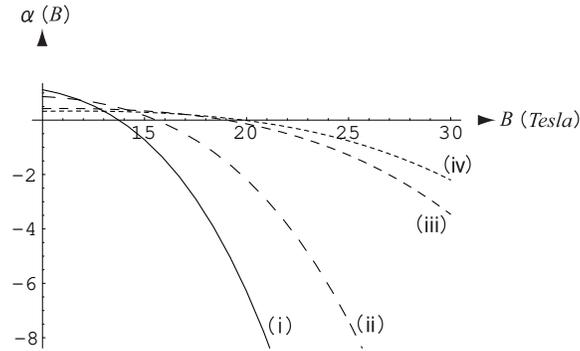}
  \end{center}
  \caption{In the case of $n=10^{23}m^{-3}$, parameter $\alpha(B)$ of interaction vs. magnetic field $B$. (i) $\epsilon=10\epsilon_0, m=0.5m_e,$ (ii) $\epsilon=11.9\epsilon_0, m=0.26m_e$, (iii) $\epsilon=13.1\epsilon_0, m=0.067m_e$ (typical values of GaAs), and (iv) $\epsilon=15\epsilon_0, m=0.05m_e$, where $\epsilon_0$ is the permittivity of vacuum and $m_e$ is the rest mass of electron.}
\label{Figalpha1}
\end{figure}

\begin{figure}[htbp]
  \begin{center}
    \includegraphics[scale=0.8]{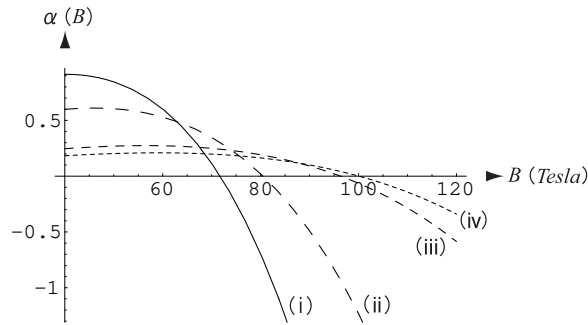}
  \end{center}
  \caption{In the case of $n=10^{24}m^{-3}$, parameter $\alpha(B)$ of interaction vs. magnetic field $B$. (i) $\epsilon=10\epsilon_0, m=0.5m_e,$ (ii) $\epsilon=11.9\epsilon_0, m=0.26m_e$, (iii) $\epsilon=13.1\epsilon_0, m=0.067m_e$ (typical values of GaAs), and (iv) $\epsilon=15\epsilon_0, m=0.05m_e$.}
\label{Figalpha2}
\end{figure}

From eq. (\ref{rgprob3}) we find that the transmission probability vanishes as $T\to 0$ if $\alpha(B)>0$, as in the case of one-dimensional systems. On the other hand, in Fig.~\ref{Figalpha1} and Fig.~\ref{Figalpha2}, we see that $\alpha(B)$ can be negative for large magnetic fields. In this case, according to eq. (\ref{rgprob3}), the transmission probability should become 1 at 0K. Since the calculation of $\alpha(B)$ is based on random phase approximation, it is not clear whether 
the values of $\alpha(B)$ are quantitatively reliable. Nevertheless, the tendency that $\alpha(B)$ decreases as the magnetic field increases should be reliable. 
As can be seem from eq. (\ref{hfpara}) the Fock term suppresses the transmission while the Hartree term enhances it. The role of the Hartree term can be interpreted as the screening of the barrier potential, and it is not surprising if the barrier becomes transparent because of the strong screening due to the one-dimensional nature of the system. 

\section{Summaries}

We investigated the effects of electron-electron interaction on the transmission of electrons through a tunnel barrier in a strong magnetic field. We have found that the temperature dependence of the transmission probability is at large the same as that of one-dimensional systems, as in Refs.~\citen{rf:16} and \citen{rf:17}. In fact, such behavior can be expected because in both cases the Friedel oscillations of the electron density play an essential role. 

In estimating the parameter $\alpha(B)$, we have shown that we should use dynamical screened coulomb interaction and bare coulomb interaction for Fock correction and Hartree correction, respectively: It is important to use approximations as good as possible, in order to verify the interaction effect from the experimental data.

It is interesting to investigate the magnetic field dependence of $\alpha(B)$. To do so, one need not go to very low temperature. In fact, at moderate temperatures from eq. (\ref{rgprob1}) we can expect that
\begin{equation}
\delta{\cal T} = 2\alpha(B) {\cal T}_0(1-{\cal T}_0)\ln(k_B T)\,.
\end{equation}
If $\alpha(B)$ decreases with the increase of the magnetic field, we can expect it to become negative  for a larger magnetic field, as in Figs.~\ref{Figalpha1} and \ref{Figalpha2}.

Some line in these figures are for the values of effective mass etc. of typical semiconductor. In the case of doped semiconductors, the scatterings of electrons by impurities will suppress the long tail of the Friedel oscillations. In order to observe clear interaction effects, the mean free path of the electrons has to be much longer than the Fermi wave length.

Very pure semimetals are good candidates for the observation of the interaction effects. In semimetals, however, the band structures are generally very complex, and the theory needs some modifications to be applied to them.

\section*{Acknowledgments}

This work is partly supported by the "High Technology Research Center Project" of Ministry of Education, Culture, Sports, Sciences and
Technology.

\begin{appendix}

\section{The derivation of eq.~(\ref{wf4}) by Green's function method}
First we consider only the Hartree term. From Eqs. (\ref{green3}) and (\ref{self1}), the corresponding part of the Matsubara Green's function ${\cal G}^H(\mib{x},\mib{y},\omega_n)$ is given by
\begin{equation}
{\cal G}^H(\mib{x},\mib{y},\omega_n)=\frac{1}{\hbar}\int d\mib{x}_1{\cal G}^0(\mib{x},\mib{x}_1,\omega_n)V_H(\mib{x}_1){\cal G}^0(\mib{x}_1,\mib{y},\omega_n),
\end{equation}
where $V_H(\mib{x}_1)$ is the Hartree-potential
\begin{equation}
V_H(\mib{x}_1)=\int d\mib{x}_2V(\mib{x}_1-\mib{x}_2)\rho(\mib{x}_2)\,,
\end{equation}
$\rho(\mib{x}_2)$ being the electron density.
Thus the corresponding retarded Green's function can be written as
\begin{equation}
G_R^H(\mib{x},\mib{y},\omega)=\frac{1}{\hbar}\int\frac{d\mib{k}'}{(2\pi)^2}\int\frac{d\mib{k}''}{(2\pi)^2}\int d\mib{x}_1\frac{\varphi_{\mib{k}'}^0(\mib{x}){\varphi_{\mib{k}'}^0}^*(\mib{x}_1)}{\omega-\xi_{\mib{k}'}^0/\hbar+i\eta}
V_H(\mib{x}_1)\frac{\varphi_{\mib{k}''}^0(\mib{x}_1){\varphi_{\mib{k}''}^0}^*(\mib{y})}{\omega-\xi_{\mib{k}''}^0/\hbar+i\eta}.\label{agreen1}
\end{equation}
The Fourier transform of eq. (\ref{agreen1}) is given by
\begin{align}
G_R^H(\mib{x},\mib{y},\tau-\tau_0)=&\frac{1}{\hbar}\int\frac{d\mib{k}'}{(2\pi)^2}\int\frac{d\mib{k}''}{(2\pi)^2}\int d\mib{x}_1\varphi_{\mib{k}'}^0(\mib{x}){\varphi_{\mib{k}'}^0}^*(\mib{x}_1)V_H(\mib{x}_1)\varphi_{\mib{k}''}^0(\mib{x}_1){\varphi_{\mib{k}''}^0}^*(\mib{y})\nonumber\\
&\times\int\frac{d\omega}{2\pi}\frac{e^{-i\omega(\tau-\tau_0)}}{(\omega-\xi_{\mib{k}'}^0/\hbar+i\eta)(\omega-\xi_{\mib{k}''}^0/\hbar+i\eta)}.
\label{A4}
\end{align}

The correction to the wave functions $\varphi_{\mib{k}}^H(\mib{x},\tau)$ by the Hartree term is given by
\begin{equation}
\varphi_{\mib{k}}^{(1H)}(\mib{x},\tau)=i\int d\mib{y}G_R^H(\mib{x},\mib{y},\tau-\tau_0)\varphi_{\mib{k}}^0(\mib{y},\tau_0)\,,
\end{equation}
where $\varphi_{\mib{k}}(\mib{y},\tau_0)=\varphi_{\mib{k}}^0(\mib{y})e^{-i\xi_{\mib{k}}^0\tau_0/\hbar}$ and we will let $\tau_0\to-\infty$. Putting eq.~(\ref{A4}) into the above expression and performing the integral over $\mib{y}$, $\mib{k}''$, and $\omega$, we obtain
\begin{align}
\varphi_{\mib{k}}^{(1H)}(\mib{x},\tau)=&\frac{1}{\hbar}\int\frac{d\mib{k}'}{(2\pi)^2}\int d\mib{x}_1\varphi_{\mib{k}'}^0(\mib{x}){\varphi_{\mib{k}'}^0}^*(\mib{x}_1)V_H(\mib{x}_1)\varphi_{\mib{k}}^0(\mib{x}_1,\tau_0)\nonumber\\
&\times\left\{\frac{e^{-i\xi_{\mib{k}}^0(\tau-\tau_0)/\hbar}}{\xi_{\mib{k}}^0/\hbar-\xi_{\mib{k}'}^0/\hbar}+\frac{e^{-i\xi_{\mib{k}'}^0(\tau-\tau_0)/\hbar}}{\xi_{\mib{k}'}^0/\hbar-\xi_{\mib{k}}^0/\hbar} \right\}.
\end{align}
In the limit $\tau_0\to-\infty$, the second term in the curly bracket of the above equation oscillates very rapidly when the integral over $\mib{k}'$ is done, hence its contribution can be neglected. Therefore we have 
\begin{align}
\varphi_{\mib{k}}^{(1H)}(\mib{x},\tau)&=e^{-i\xi_{\mib{k}}^0\tau/\hbar}\int d\mib{x}_1\int\frac{d\mib{k}'}{(2\pi)^2}\frac{\varphi_{\mib{k}'}^0(\mib{x}){\varphi_{\mib{k}'}^0}^*(\mib{x}_1)}{\xi_{\mib{k}}^0-\xi_{\mib{k}'}^0} V_H(\mib{x}_1)\varphi_{\mib{k}}^0(\mib{x}_1).
\end{align}
Hereafter we omit the time-dependence of the wave functions, and it can be written in the form
\begin{equation}
\varphi_{\mib{k}}^{(1H)}(\mib{x})=\int d\mib{x}_1G_{\mib{k}}(\mib{x};\mib{x}_1)V_H(\mib{x}_1)\varphi_{\mib{k}}^0(\mib{x}_1),\label{awf1}
\end{equation}
where $G_{\mib{k}}(\mib{x};\mib{x}_1)$ is the single-electron Green's function for noninteracting electrons and is defined as
\begin{equation}
G_{\mib{k}}(\mib{x};\mib{x}_1)\equiv\int\frac{d\mib{k}'}{(2\pi)^2}\frac{\varphi_{\mib{k}'}^0(\mib{x}){\varphi_{\mib{k}'}^0}^*(\mib{x}_1)}{\xi_{\mib{k}}^0-\xi_{\mib{k}'}^0}.\label{agreen2}
\end{equation}
As for the treatment of the singularity of the integrand, the reader is referred to the Appendix B.

Next we consider the Fock term. Using Eqs. (\ref{green3}) and (\ref{self1}), we can write the corresponding part of the Matsubara Green's function ${\cal G}^F(\mib{x},\mib{y},\omega_n)$ as
\begin{equation}
{\cal G}^F(\mib{x},\mib{y},\omega_n)=-\frac{1}{\hbar}\int\int d\mib{x}_1d\mib{x}_1'{\cal G}^0(\mib{x},\mib{x}_1,\omega_n)V_F(\mib{x}_1,\mib{x}_1'){\cal G}^0(\mib{x}_1',\mib{y},\omega_n)
\end{equation}
where $V_F(\mib{x}_1,\mib{x}_1')$ is the Fock-potential
\begin{equation}
V_F(\mib{x}_1,\mib{x}_1')=V(\mib{x}_1-\mib{x}_1')\int\frac{d\mib{k}'}{(2\pi)^2}\varphi_{\mib{k}'}^0(\mib{x}_1){\varphi_{\mib{k}'}^0}^*(\mib{x}_1').
\end{equation}
Thus the corresponding retarded Green's function can be written as
\begin{align}
G_{R}^{F}(\mib{x},\mib{y},\omega)=&-\frac{1}{\hbar}\int\frac{d\mib{k}'}{(2\pi)^2}\int\frac{d\mib{k}''}{(2\pi)^2}\iint d\mib{x}_1d\mib{x}_1'\nonumber\\
&\times\frac{\varphi_{\mib{k}'}^0(\mib{x}){\varphi_{\mib{k}'}^0}^*(\mib{x}_1)}{\omega-\xi_{\mib{k}'}^0/\hbar+i\eta}V_F(\mib{x}_1,\mib{x}_1')\frac{\varphi_{\mib{k}''}^0(\mib{x}_1){\varphi_{\mib{k}''}^0}^*(\mib{y})}{\omega-\xi_{\mib{k}''}^0/\hbar+i\eta}.\label{agreen3}
\end{align}
The Fourier transform of eq. (\ref{agreen3}) is given by
\begin{align}
G_{R}^{F}(\mib{x},\mib{y},\tau-\tau_0)&=-\frac{1}{\hbar}\int\frac{d\mib{k}'}{(2\pi)^2}\int\frac{d\mib{k}''}{(2\pi)^2}\iint d\mib{x}_1d\mib{x}_1'\varphi_{\mib{k}'}^0(\mib{x}){\varphi_{\mib{k}'}^0}^*(\mib{x}_1)V_F(\mib{x}_1,\mib{x}_1')\nonumber\\
& \times\varphi_{\mib{k}''}^0(\mib{x}_1'){\varphi_{\mib{k}''}^0}^*(\mib{y})\int\frac{d\omega}{2\pi}\frac{e^{-i\omega(\tau-\tau_0)}}{(\omega-\xi_{\mib{k}'}^0/\hbar+i\eta)(\omega-\xi_{\mib{k}''}^0/\hbar+i\eta)}.
\end{align}
Thus, as in the case of Hartree term, we can calculate the correction to $\varphi_{\mib{k}}^F(\mib{x},\tau)$ by the Fock term from
\begin{equation}
\varphi_{\mib{k}}^{(1F)}(\mib{x},\tau)=i\int d\mib{y}G_R^F(\mib{x},\mib{y},\tau-\tau_0)\varphi_{\mib{k}}(\mib{y},\tau_0)\,,
\end{equation}
i.e.,
\begin{equation}
\varphi_{\mib{k}}^{(1F)}(\mib{x})=-\iint d\mib{x}_1d\mib{x}_1'G_{\mib{k}}(\mib{x};\mib{x}_1)V_F(\mib{x}_1,\mib{x}_1')\varphi_{\mib{k}}^0(\mib{x}_1'),\label{awf2}
\end{equation}
where we omit the time-dependence of the wave functions and $G_{\mib{k}}(\mib{x};\mib{x}_1)$ is defined by eq. (\ref{agreen2}). From Eqs. (\ref{awf1}) and (\ref{awf2}), we obtain eq. (\ref{wf4}). 

\section{The calculation of the single-electron Green's function}

To calculate the correction to the transmission probability we need only the asymptotic form of the $G_{\mib{k}}(\mib{x},\mib{x}_1)$ at $z\to+\infty$. From eqs. (\ref{green4}), (\ref{wf0}) and (\ref{ukt}) we have
\begin{align}
G_{\mib{k}}(\mib{x},\mib{x}_1)&=\int\frac{dk_y'}{2\pi}\phi_{k_y'}^0(x,y){\phi_{k_y'}^0}^*(x_1,y_1)\cdot\int\frac{dk_z'}{2\pi}\frac{u_{k_z'}^0(z){u_{k_z'}^0}^*(z_1)}{\xi_{\mib{k}}^0-\xi_{\mib{k}'}^0}\nonumber\\
& \equiv G_{k_y}(x,y;x_1,y_1)\cdot G_{k_z}(z;z_1),
\label{eb1}
\end{align}
Note that the energy eigenvalues $\xi_{\mib{k}}^0$ depend only on $k_z$ and first we calculate $G_{k_y}(x,y;x_1,y_1)$:
\begin{align}
G_{k_y}(x,y;x_1,y_1)&\equiv\int\frac{dk_y'}{2\pi}\phi_{k_y'}^0(x,y){\phi_{k_y'}^0}^*(x_1,y_1)\nonumber\\
&= \frac{1}{\pi^{1/2}\ell_B}\int_{-L_x/2{\ell_B}^2}^{L_x/2{\ell_B}^2}\frac{dk_y'}{2\pi}e^{ik_y'(y-y_1)}\nonumber\\
&\times \exp\left[-\frac{(x+k_y'{\ell_B}^2)^2 +
(x_1+k_y'{\ell_B}^2)^2}{2{\ell_B}^2}\right].
\end{align}
We let $L_x\to\infty$, then we obtain
\begin{align}
G_{k_y}(x,y;x_1,y_1)&=\frac{1}{2\pi{\ell_B}^2}\exp\left[-\frac{(x-x_1)^2+(y-y_1)^2}{4{\ell_B}^2}\right]\nonumber\\
&\times\exp\left[-\frac{i(x+x_1)(y-y_1)}{2{\ell_B}^2} \right].\label{b3}
\end{align}

\begin{figure}[htbp]
  \begin{center}
    \includegraphics[scale=0.5]{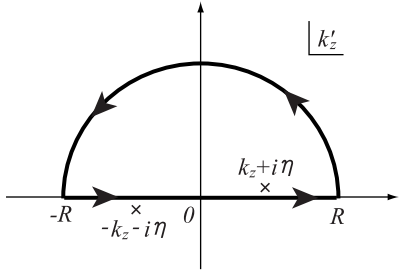}
  \end{center}
  \caption{The contour of the first integral in eq.~(\ref{B6}) ($R \to \infty$).}
\label{Fig.7}
\end{figure}

Next we calculate $G_{k_z}(z;z_1)$ for $k_z > 0$: 
\begin{equation}
G_{k_z}(z;z_1) \equiv \int\frac{dk_z'}{2\pi}\frac{u_{k_z'}^0(z){u_{k_z'}^0}^*(z_1)}{\xi_{\mib{k}}^0-\xi_{\mib{k}'}^0}\,.
\label{b5}
\end{equation}
In treating the singularities of the integral, we replace $k_z$ with $k_z + i \eta\, (\eta \to + 0)$ so that we extract only the out going waves at $z\to+\infty$.
Then, for $z_1 > 0$, from eqs.~(\ref{uk1}) and (\ref{uk2}) we have
\begin{align}
G_{k_z}(z;z_1)&=\int_{0}^{\infty}\frac{dk_z'}{2\pi}\frac{|t_0|^2 e^{ik_z'(z - z_1)}}{\hbar v_F (k_z + i\eta - k_z')}\nonumber\\
&+\int_{-\infty}^0 \frac{dk_z'}{2\pi}\frac{(e^{ik_z'z}+r_0 e^{-ik_z'z})(e^{-ik_z'z_1}+{r_0}^* e^{ik_z'z_1})}{\hbar v_F (k_z + i\eta+ k_z')},
\label{B6}
\end{align}
where we used the linearized energy dispersion relation eq.~(\ref{dispersion}).
Since the integrands of the above integrals oscillate very rapidly for large $z$, the main contributions come from only such $k_z'$ that the denominators vanish. Therefore we may extend the regions of the integrals to $(-\infty, \infty)$ and, after adding a contour of an infinitely large semi-circle, we obtain (see Fig.~\ref{Fig.7}) 
\begin{equation}
G_{k_z}(z;z_1) = \frac{1}{i\hbar v_F}\left\{e^{i k_z(z - z_1)} + r_0 e^{i k_z(z + z_1)} \right\}\,.
\label{b1}
\end{equation}

Similarly, for $z_1<0$, we obtain
\begin{equation}
G_{k_z}(z;z_1)=\frac{1}{i\hbar v_F}t_0e^{ik_z(z-z_1)},\label{b2}
\end{equation}
where we used the relation
\begin{equation}
t_0{r_0}^*+{t_0}^*r_0=0\,,
\end{equation}
which can be derived from a time reversal symmetry argument.

Thus from eqs.~(\ref{eb1}), (\ref{b3}), (\ref{b1}) and (\ref{b2}) we obtain eq.~(\ref{green5}).

\section{The calculations of the correction to the transmission amplitude by the perturbation theory}

\subsection{The derivation of eq. (\ref{fwf2})}

In eq. (\ref{fwf2}), because of $V(\mib{x}_1-\mib{x}_1')$, main contribution comes from the regions where $z_1$ and $z_1'$ are of the same sign. First we consider only the integration over $k_z'$ in the Fock potential. When both $z_1$ and $z_1'$ are positive, it is easily found that
\begin{align}
& \int_{-k_F}^{k_F}\frac{dk_z'}{2\pi}u_{k_z'}^0(z_1){u_{k_z'}^0}^*(z_1')\nonumber\\
&=|t_0|^2\int_{0}^{k_F}\frac{dk_z'}{2\pi}e^{ik_z'(z_1-z_1')}+|r_0|^2\int_{-k_F}^{0}\frac{dk_z'}{2\pi}e^{-ik_z'(z_1-z_1')}\nonumber\\
&+\int_{-k_F}^{0}\frac{dk_z'}{2\pi}e^{ik_z'(z_1-z_1')}+\int_{-k_F}^{0}\frac{dk_z'}{2\pi}\left[{r_0}^*e^{ik_z'(z_1+z_1')}+c.c. \right]\nonumber\\
&=\frac{\sin[k_F(z_1-z_1')]}{\pi(z_1-z_1')}\nonumber\\
&+|r_0|\frac{\sin[k_F(z_1+z_1')+\arg r_0]-\sin(\arg r_0)}{\pi(z_1+z_1')}.
\end{align}
Similarly, when both $z_1$ and $z_1'$ are negative, it is easily found that
\begin{align}
& \int_{-k_F}^{k_F}\frac{dk_z'}{2\pi}u_{k_z'}^0(z_1){u_{k_z'}^0}^*(z_1')\nonumber\\
&=\frac{\sin[k_F(z_1-z_1')]}{\pi(z_1-z_1')}+|r_0|\frac{\sin[k_F(z_1+z_1')-\arg r_0]+\sin(\arg r_0)}{\pi(z_1+z_1')}.
\end{align}
Next we transform $z_1$ and $z_1'$ into $z_2=z_1+z_1'$ and $z_2'=z_1-z_1'$, respectively. Then the left side of eq. (\ref{fwf2}) becomes
\begin{align}
& \frac{t_0r_0|r_0|}{2\pi i\hbar v_F}e^{ik_zz}\int_{-\infty}^{\infty} dz_2'V(x_1-x_1',y_1-y_1',z_2')\nonumber\\
&\times\left[\int_{0}^{\infty}dz_2\frac{e^{ik_z z_2}}{z_2}\sin(k_F z_2+\arg r_0)+\int_{-\infty}^{0}dz_2\frac{e^{-ik_z z_2}}{z_2}\sin(k_F z_2-\arg r_0)\right]\nonumber\\
&=\frac{t_0r_0|r_0|}{\pi i\hbar v_F}e^{ik_zz}\tilde{V}(x_1-x_1',y_1-y_1';0)\int_{0}^{\infty}dz_2\frac{e^{ik_z z_2}}{z_2}\sin(k_Fz_2+\arg r_0),\label{c1}
\end{align}
where $\tilde{V}(x_1-x_1',y_1-y_1';k_z)$ is the Fourier transform for only $z$ component and defined by eq.~(\ref{c6})

Since the expression for the electron density, eq. (\ref{density}), is only valid at large distances $|z|$, we restrict the integration in eq. (\ref{c1}) in the regions $|z|\ge d$. At low temperatures, the electronic conduction is determined entirely by electrons with energy close to the Fermi energy. Therefore in the integral in eq.~(\ref{c1}), we leave only the terms which are divergent for $k_z \to k_F$. Then, we find that the integral becomes
\begin{equation}
\frac{i}{2}e^{-i\arg r_0}\int_d^{\infty}dz_2\frac{\cos[(k_z-k_F)z_2]}{z_2},
\end{equation}
and that for $|k_z-k_F|d \ll 1 $ eq. (\ref{c1}) reduces to
\begin{equation}
\frac{t_0|r_0|^2}{2\pi\hbar v_F}\tilde{V}(x_1-x_1',y_1-y_1';0)e^{ik_zz}\ln\left(\frac{1}{|k_z-k_F|d} \right).
\label{c2}
\end{equation}

\subsection{The derivation of eq. (\ref{fwf4})}

The integral over $k_y'$ in eq. (\ref{fwf3}) is easily found to be
\begin{align}
&\int\frac{dk_y'}{2\pi}\phi_{k_y'}^0(x_1,y_1){\phi_{k_y'}^0}^*(x_1',y_1')\nonumber\\
=&\frac{1}{2\pi{\ell_B}^2}\exp\left[-\frac{(x_1-x_1')^2+(y_1-y_1')^2}{4{\ell_B}^2} \right]\exp\left[-\frac{i(x_1+x_1')(y_1-y_1')}{2{\ell_B}^2} \right].
\end{align}
Then we find that eq. (\ref{fwf3}) is given by
\begin{align}
\varphi_{\mib{k}}^{(1F)}(\mib{x})=&-\frac{t_0|r_0|^2}{2\pi\hbar v_F}\ln\left(\frac{1}{|k_z-k_F|d} \right)\cdot e^{ik_zz}\iiiint_{-\infty}^{\infty}dx_1dy_1dx_1'dy_1'\nonumber\\
&\times\left\{\frac{1}{2\pi{\ell_B}^2}\exp\left[-\frac{(x_1-x_1)^2+(y_1-y_1)^2}{4{\ell_B}^2} \right]\exp\left[-\frac{i(x_1+x_1)(y_1-y_1)}{2{\ell_B}^2} \right]\right\}\nonumber\\
&\times\left\{\frac{1}{2\pi{\ell_B}^2}\exp\left[-\frac{(x_1-x_1')^2+(y_1-y_1')^2}{4{\ell_B}^2} \right]\exp\left[-\frac{i(x_1+x_1')(y_1-y_1')}{2{\ell_B}^2} \right]\right\}\nonumber\\
&\times\frac{1}{\pi^{1/4}{\ell_B}^{1/2}}\exp\left[-\frac{(x_1'+k_y{\ell_B}^2)^2 }{2{\ell_B}^2}\right]e^{ik_yy_1'}\tilde{V}_s(x_1-x_1',y_1-y_1';0)\,,
\end{align}
and after performing the integrals, we obtain eq.~(\ref{fwf4}).

\subsection{The derivation of eq. (\ref{hwf2})}

From eq. (\ref{b1}) and (\ref{b2}), we have
\begin{align}
&\int_{-\infty}^{\infty}dz_1G_{k_z}(z;z_1)\left[\int_{|z_2|\ge d} dz_2V(\mib{x}_1-\mib{x}_2)n(z_2) \right]u_{k_z}^0(z_1)\nonumber\\
&=\frac{t_0}{i\hbar v_F} e^{ik_zz}\int_{-\infty}^{\infty}dz_1\int_{|z_2|\ge d} dz_2V(\mib{x}_1-\mib{x}_2)n(z_2)\nonumber\\
&+\frac{t_0r_0}{i\hbar v_F}e^{ik_zz}\left(\int_{-\infty}^{0}dz_1e^{-2ik_zz_1}+
\int_{0}^{\infty}dz_1e^{2ik_zz_1} \right)\int_{|z_2|\ge d} dz_2V(\mib{x}_1-\mib{x}_2)n(z_2)\,,
\label{c4}
\end{align}
where the integrations over $z_2$ are done for $|z_2|\ge d$ because the expression for the electron density $n(z_2)$ is valid only at large distances $|z_2|$. In eq. (\ref{c4}), we neglect the first term of the right hand side because its integrand oscillates and its contribution to the integral is very small. Then the right hand side of eq. (\ref{c4}) is easily found to be
\begin{equation}
\frac{t_0r_0}{i\hbar v_F}e^{ik_zz}\left[\int_{-\infty}^{\infty}dz_1
\cos(2k_zz_1)+2i\int_{0}^{\infty}dz_1\sin(2k_zz_1) \right]\int_{|z_2|\ge d}dz_2V(\mib{x}_1-\mib{x}_2)n(z_2)\,.
\label{c5}
\end{equation}
Here we consider the integral with $\cos(2k_zz_1)$:
\begin{equation}
I_1 \equiv \int_{-\infty}^{\infty}dz_1\cos(2k_zz_1)\int_{|z_2|\ge d}
dz_2 V(\mib{x}_1-\mib{x}_2)n(z_2)\,.
\end{equation}
Changing the integration over $z_1$  to that over  $z_1'\equiv z_1-z_2$ and making use of that $V(\mib{x}_1-\mib{x}_2)$ is an even function of the arguments, we easily find that
\begin{equation}
I_1 = \tilde{V}(x_1-x_2,y_1-y_2;2k_z)\int_{|z_2|\ge d}dz_2 \cos (2k_z z_2) n(z_2)\,,
\end{equation}
where we used  eq. (\ref{c6}).

In the integral over $z_2$, we leave only the term which is divergent at $k_z \to k_F$. Then from eq.~(\ref{density}) we easily find that
\begin{align}
I_1 &= -\frac{|r_0|}{2\pi}\tilde{V}(x_1-x_2,y_1-y_2;2k_z)\int_d^{\infty}\frac{dz_2}{z_2}\sin[2(k_z-k_F)z_2-\arg r_0]\nonumber\\
&=\frac{|r_0|}{2\pi}\tilde{V}(x_1-x_2,y_1-y_2;2k_z)\sin(\arg r_0)
\ln\left(\frac{1}{|k_z-k_F|d} \right).
\end{align}
In the same way, we can calculate the other term in eq.~(\ref{c5}) and we find that the right side of eq.~(\ref{c4}) becomes
\begin{equation}
\frac{t_0|r_0|^2}{2\pi\hbar v_F}e^{ik_zz}\tilde{V}(x_1-x_2,y_1-y_2;2k_F)\ln\left(\frac{1}{|k_z-k_F|d} \right),
\label{c7}
\end{equation}
where we have replaced $2k_z$ with $2k_F$ in the potential.
\subsection{The derivation of eq. (\ref{hwf4})}

We easily find that $\tilde{V}(x_1-x_2,y_1-y_2;2k_F)$ in eq. (\ref{hwf2}) is given by
\begin{equation}
\tilde{V}(x_1-x_2,y_1-y_2;2k_F)=\frac{e^2}{2\pi\epsilon}K_0\left[\left\{(x_1-x_2)^2+(y_1-y_2)^2 \right\}^{1/2}\cdot2{k_F} \right]\,,
\end{equation}
where $K_0(r)$ is the modified Bessel function.
Then from eqs. (\ref{b3}) and (\ref{c7}), we find that eq. (\ref{hwf2}) becomes
\begin{align}
\varphi_{\mib{k}}^{(1H)}(\mib{x})&=\frac{t_0|r_0|^2}{2\pi\hbar v_F}\frac{e^2}{2\pi\epsilon}\ln\left(\frac{1}{|k_z-k_F|d} \right)\cdot e^{ik_zz}\nonumber\\
&\times\iiiint_{-\infty}^{\infty}dx_1dy_1dx_2dy_2\left\{\frac{1}{2\pi{\ell_B}^2}\exp\left[-\frac{1}{4{\ell_B}^2}(x-x_1)^2 \right] \right.\nonumber\\
&\times\left.\exp\left[-\frac{1}{4{\ell_B}^2}(y-y_1)^2 \right]\exp\left[-\frac{1}{2{\ell_B}^2}(x+x_1)(y-y_1) \right] \right\}\nonumber\\
&\times\frac{1}{\pi^{1/4}{\ell_B}^{1/2}}\exp\left[-\frac{1}{2{\ell_B}^2}(x_1+k_y{\ell_B}^2)^2 \right]e^{ik_yy_1}\nonumber\\
&\times\frac{1}{2\pi{\ell_B}^2}K_0\left[\left\{(x_1-x_2)^2+(y_1-y_2)^2 \right\}^{1/2}\cdot 2k_F \right].
\end{align}
We first do the integrations over $x_2$ and $y_2$, and using the formula
\begin{equation}
\int_0^{\infty}rK_0(r)dr =1\,,
\end{equation}
we have
\begin{align}
\varphi_{\mib{k}}^{(1H)}(\mib{x})&=\frac{t_0|r_0|^2}{2\pi\hbar v_F}\frac{e^2}{2\pi\epsilon}\ln\left(\frac{1}{|k_z-k_F|d} \right)\phi_{k_y}^0(x,y)\frac{1}{4{k_F}^2{\ell_B}^2}e^{ik_zz}\nonumber\\
&=\alpha_1(B)t_0(1-|t_0|^2)\ln\left(\frac{1}{|k_z-k_F|d} \right)\phi_{k_y}^0(x,y)e^{ik_zz},
\end{align}
where
\begin{align}
\alpha_1(B)&\equiv\frac{1}{2\pi\hbar v_F}\frac{e^2}{2\pi\epsilon}\frac{1}{4{k_F}^2{\ell_B}^2}\nonumber\\
&=\frac{\kappa^2}{4{k_F}^2}.
\end{align}

\section{The Screened Coulomb Interaction by RPA}
Below we will consider the case of zero-temperature, and we will not take into account the effects of the barrier on the screening, for simplicity. Suppose we put an external charge
\begin{equation}
Q(\mib{x},t)=Q_{\mib{q}}e^{i(\mib{q}\cdot\mib{x}-\omega t)}
\end{equation}
in the system, where $\mib{q}$ is a three-dimensional vector (below $\mib{k}$ will be a two-dimensional vector in $y$-$z$ plane, and we define $\mib{k} \pm \mib{q} \equiv (k_y \pm q_y, k_z \pm q_z$ )). Then the electron density induced by this external charge is given by
\begin{equation}
Q'(\mib{x},t)=\chi(\mib{q},\omega)Q(\mib{x},t).
\end{equation}
Here $\chi(\mib{q},\omega)$ is the density response function defined by \cite{rf:15}
\begin{equation}
\chi(\mib{q},\omega)=\frac{1}{i\hbar}\hat{V}(\mib{q})\int_{0}^{\infty}dt\left\langle [\rho_{-\mib{q}},\rho_{\mib{q}}(t)] \right\rangle e^{i\omega t-\eta t},
\end{equation}
where $\langle\cdots\rangle$ indicates the thermal average, $\hat{V}(\mib{q})$ is the Fourier component of the coulomb interaction potential
\begin{equation}
\hat{V}(\mib{q})=\frac{e^2}{\epsilon q^2},
\end{equation}
$\epsilon$ being the dielectric permittivity of the matter.
and $\rho_{\mib{q}}$ is the electron density operator
\begin{align}
\rho_{\mib{q}}&=\int d\mib{x}\hat{\psi}^{\dagger}(\mib{x})\hat{\psi}(\mib{x})e^{-i\mib{q}\cdot\mib{x}}\nonumber\\
&=\sum_{\mib{k}}{a_{\mib{k}-\mib{q}}}^{\dagger}a_{\mib{k}}\exp\left[i\frac{{\ell_B}^2}{2}q_x(2k_y-q_y) -\frac{{\ell_B}^2}{4}({q_x}^2 + {q_y}^2) \right].\label{d5}
\end{align}
We define the Matsubara Green's function ${\cal D}(\mib{q},\omega_n)$,
\begin{equation}
{\cal D}(\mib{q},\omega_n)=\int_{0}^{\beta\hbar}d\tau\left\langle\rho_{\mib{q}}(\tau)\rho_{-\mib{q}}\right\rangle e^{i\omega_n\tau},
\end{equation}
where $\omega_n=\pi k_BT(2n+1)/\hbar$, $n$ being an integer. Then $\chi(\mib{q},\omega)$ can be expressed in terms of ${\cal D}(\mib{q},\omega_n)$ by analytic continuation
\begin{equation}
\chi(\mib{q},\omega)=\frac{\hat{V}(\mib{q})}{\hbar}{\cal D}(\mib{q},-i\omega+\eta)\,,
\label{chiqom}
\end{equation}
and the dielectric function is defined as
\begin{equation}
\varepsilon(\mib{q},\omega)=\frac{\epsilon Q(\mib{x},t)}{Q'(\mib{x},t)+Q(\mib{x},t)}=\frac{\epsilon}{1+\chi (\mib{q},\omega)}\,.
\label{epsqom}
\end{equation}

From eq. (\ref{d5}), we find
\begin{align}
{\cal D}(\mib{q},\omega_n)&=\exp\left[-\frac{{\ell_B}^2}{2}({q_x}^2+{q_y}^2) \right]\nonumber\\
&\times\sum_{\mib{k}_1,\mib{k}_2}\int_{0}^{\beta\hbar}d\tau\left\langle {a_{\mib{k}_1-\mib{q}}}^{\dagger}(\tau)a_{\mib{k}_1}(\tau){a_{\mib{k}_2+\mib{q}}}^{\dagger}a_{\mib{k}_2}\right\rangle e^{i\omega_n\tau}\exp\left[i{\ell_B}^2q_x(k_{1y}-k_{2y}-q_y) \right].
\end{align}

\begin{figure}[htbp]
  \begin{center}
    \includegraphics[scale=0.4]{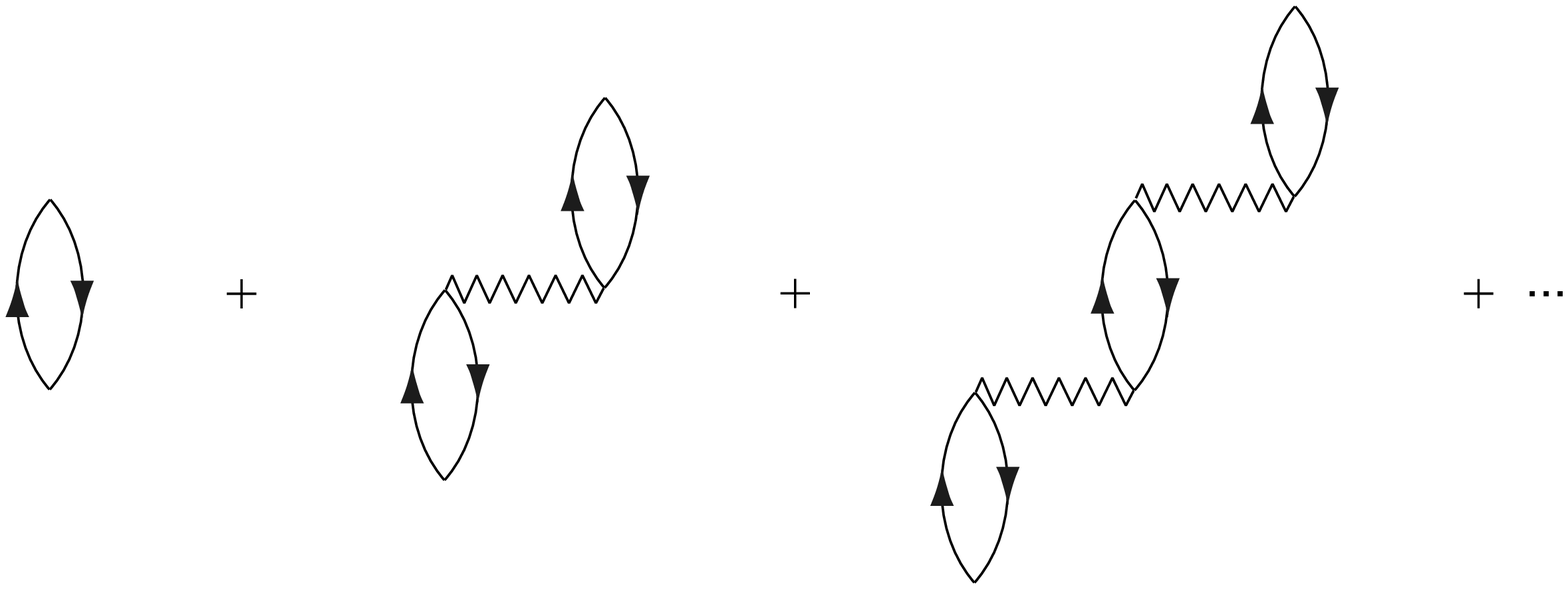}
  \end{center}
  \caption{Feynman diagram for ${\cal D}(\mib{q},\omega_n)$.}
  \label{Fig8}
\end{figure}
In calculating ${\cal D}(\mib{q},\omega_n)$ using Feynman diagram, among the various terms we retain only those corresponding to the diagrams shown in Fig. \ref{Fig8}. Then we obtain
\begin{equation}
{\cal D}(\mib{q},\omega_n)=\frac{{\cal D}^0(\mib{q},\omega_n)}{1-\hat{V}(\mib{q}){\cal D}^0(\mib{q},\omega_n)},
\label{Dqom}
\end{equation}
where
\begin{align}
{\cal D}^0(\mib{q},\omega_n)&=\exp\left[-\frac{{\ell_B}^2}{2}({q_x}^2+{q_y}^2) \right]\nonumber\\
&\times\sum_{\mib{k}}\frac{1}{\beta\hbar}\sum_{\nu_n}{\cal F}^0(\mib{k},\nu_n){\cal F}^0(\mib{k}-\mib{q},\omega_n+\nu_n),
\end{align}
with $\nu_n=\pi k_BT(2n+1)/\hbar$, $n$ being an integer and ${\cal F}^0(\mib{k},\nu_n)$ is the Matsubara Green's function for free electrons
\begin{equation}
{\cal F}^0(\mib{k},\nu_n)=\frac{1}{i\nu_n-\xi_{\mib{k}}/\hbar}.
\end{equation}
It is easy to calculate ${\cal D}^0(\mib{q},\omega_n)$:
\begin{equation}
{\cal D}^0(\mib{q},\omega_n) =e^{-l_B^2 q_{\perp}^2/2}\sum_{\mib{k}}\frac{f_F(\xi_{\mib{k}})-f_F(\xi_{\mib{k}-\mib{q}})}{(\xi_{\mib{k}}-\xi_{\mib{k}-\mib{q}})/\hbar+i\omega_n}\,,
\label{D0qomn1}
\end{equation}
where $f_F(\xi_{\mib{p}})$ is the Fermi distribution function and $q_{\perp}^2 = q_x^2 + q_y^2$\,.

From eqs.~(\ref{chiqom}), (\ref{epsqom}) and (\ref{Dqom}), the analytic continuation of the dielectric function is given by
\begin{equation}
\varepsilon(\mib{q},\omega_n)=\epsilon\left[1-\hat{V}(\mib{q}){\cal D}^0(\mib{q},\omega_n) \right].
\label{dielectric}
\end{equation}
The screened interaction is given by
\begin{equation}
\hat{V}_s(\mib{q},\omega_n)=\frac{e^2}{q^2\varepsilon(\mib{q},\omega_n)}\,.
\label{epsqom2}
\end{equation}

As is mentioned in the last part of subsection \ref{subsecFock}, the relevant parameter to the Fock correction is given by eq.~(\ref{limVqom}).
For small $q_z$, from eq.~(\ref{D0qomn1}) we have
\begin{equation}
{\cal D}^0(\mib{q},\omega_n) = - \frac{e^{-l_B^2 q_{\perp}^2/2}}{4\pi^2 l_B^2 \hbar v_F}\frac{2 (v_F q_z)^2}{( v_F q_z)^2 + \omega_n^2}\,,
\label{D0qomn}
\end{equation}
hence we find that
\begin{equation}
\lim_{q_z \to 0}\hat{V}_s(\mib{q}, v_F q_z) = \frac{e^2}{\epsilon\left\{q_{\perp}^2+\kappa^2\exp\left[-\frac{1}{2}(\ell_B\, q_{\perp})^2 \right] \right\}}\,,
\end{equation}
where $\mib{q}_{\perp}\equiv (q_x,  q_y, 0)$ and
\begin{equation}
\kappa\equiv\sqrt{\frac{e^2}{4\pi^2\epsilon\hbar v_F{\ell_B}^2}}\,.
\label{kappa}
\end{equation}

\section{The derivation of eq. (\ref{rgprob1})}
First we consider only the Hartree term. The 1st-order Hartree correction to the transmitted wave with cutoff $\Lambda_0$ can be written as
\begin{align}
\varphi_{\mib{k}}^{(1H)}(\mib{x}, \Lambda_0)=&\int d\mib{x}_1G_{\mib{k}}(\mib{x};\mib{x}_1)V_H(\mib{x}_1)\varphi_{\mib{k}}^0(\mib{x}_1)\nonumber\\
=&\iint dx_1dy_1G_{k_y}(x,y;x_1,y_1)\phi_{k_y}^0(x_1,y_1)\iint \frac{dx_2dy_2}{2\pi{\ell_B}^2}u_{k_z}^{1H}(\mib{x}',\Lambda_0),\label{e1}
\end{align}
where $\mib{x}'=(x_1-x_2,y_1-y_2,z)$ and
\begin{equation}
u_{k_z}^{1H}(\mib{x}',\Lambda_0)\equiv\int dz_1G_{k_z}(z;z_1)\left[\int dz_2V_s(\mib{x}_1-\mib{x}_2)n(z_2) \right]u_{k_z}^0(z_1).
\end{equation}
Here only $u_{k_z}^{1H}(\mib{x}',\Lambda_0)$ is dependent on the cutoff.
Within the strip shown in Fig.~\ref{PoorRGT}, using linearized energy dispersion, the single-electron Green's function for noninteracting electrons can be written as (see eq. (\ref{b5}))
\begin{align}
G_{k_z}(z;z_1) & =\int_{k_F-\Lambda_0}^{k_F+\Lambda_0}\frac{dk_z'}{2\pi}\frac{u_{k_z'}^0(z){u_{k_z'}^0}^*(z_1)}{\xi_{\mib{k}}^0-\xi_{\mib{k}'}^0}\nonumber\\
&=\frac{1}{2\pi\hbar v_F}\int_{-\Lambda_0}^{-s\Lambda_0}d\tilde{k}_z'\frac{u_{\tilde{k}_z'+k_F}^0(z){u_{\tilde{k}_z'+k_F}^0}^*(z_1)}{k_z-\tilde{k}_z'-k_F}\nonumber\\
&+\frac{1}{2\pi\hbar v_F}\int_{-s\Lambda_0}^{s\Lambda_0}d\tilde{k}_z'\frac{u_{\tilde{k}_z'+k_F}^0(z){u_{\tilde{k}_z'+k_F}^0}^*(z_1)}{k_z-\tilde{k}_z'-k_F}\nonumber\\
&+\frac{1}{2\pi\hbar v_F}\int_{s\Lambda_0}^{\Lambda_0}d\tilde{k}_z'\frac{u_{\tilde{k}_z'+k_F}^0(z){u_{\tilde{k}_z'+k_F}^0}^*(z_1)}{k_z-\tilde{k}_z'-k_F}\nonumber\\
&\equiv  G_{k_z}^{(A)}(z;z_1)+G_{k_z}^{(B)}(z;z_1)+G_{k_z}^{(C)}(z;z_1),
\end{align} 
where $(A)$, $(B)$ and $(C)$ denote the regions of $-\Lambda_0\le \tilde{k}_z'\le -s\Lambda_0$, $-s\Lambda_0\le \tilde{k}_z'\le s\Lambda_0$ and $s\Lambda_0\le \tilde{k}_z'\le \Lambda_0$, respectively. Similarly, the electron density $n(z_2)$ can be written as
\begin{align}
n(z_2)&=\int_{k_F-\Lambda_0}^{k_F}\frac{dk_z}{2\pi}|u_{k_z}^0(z_2)|^2\nonumber\\
&=\int_{-\Lambda_0}^{-s\Lambda_0}\frac{d\tilde{k}_z}{2\pi}|u_{\tilde{k}_z+k_F}^0(z_2)|^2+\int_{-s\Lambda_0}^{0}\frac{d\tilde{k}_z}{2\pi}|u_{\tilde{k}_z+k_F}^0(z_2)|^2\nonumber\\
&\equiv n^{(A)}(z_2)+n^{(B)}(z_2).
\end{align}
Thus $u_{k_z}^{(1H)}(\mib{x}',\Lambda_0)$ is of the form
\begin{align}
u_{k_z}^{(1H)}(\mib{x}',\Lambda_0)=&\int dz_1G_{k_z}^{(A)}(z;z_1)\left[\int dz_2V(\mib{x}_1-\mib{x}_2)n^{(A)}(z_2) \right]u_{k_z}^0(z_1)\nonumber\\
&+\int dz_1G_{k_z}^{(A)}(z;z_1)\left[\int dz_2V(\mib{x}_1-\mib{x}_2)n^{(B)}(z_2) \right]u_{k_z}^0(z_1)\nonumber\\
&+\int dz_1G_{k_z}^{(B)}(z;z_1)\left[\int dz_2V(\mib{x}_1-\mib{x}_2)n^{(A)}(z_2) \right]u_{k_z}^0(z_1)\nonumber\\
&+\int dz_1G_{k_z}^{(B)}(z;z_1)\left[\int dz_2V(\mib{x}_1-\mib{x}_2)n^{(B)}(z_2) \right]u_{k_z}^0(z_1)\nonumber\\
&+\int dz_1G_{k_z}^{(C)}(z;z_1)\left[\int dz_2V(\mib{x}_1-\mib{x}_2)n^{(A)}(z_2) \right]u_{k_z}^0(z_1)\nonumber\\
&+\int dz_1G_{k_z}^{(C)}(z;z_1)\left[\int dz_2V(\mib{x}_1-\mib{x}_2)n^{(B)}(z_2) \right]u_{k_z}^0(z_1).\label{e2}
\end{align}
We find that the 4th term of the right hand side in eq. (\ref{e2}) can be written as $u_{k_z}^{(1H)}(\mib{x}_0,s\Lambda_0)$,  hence
\begin{equation}
u_{k_z}^{(1H)}(\mib{x}',\Lambda_0)=u_{k_z}^{(1H)}(\mib{x}',s\Lambda_0)+\delta u_{k_z}^{(1H)}(\mib{x}',\Lambda_0),
\end{equation}
where
\begin{align}
\delta u_{k_z}^{(1H)}(\mib{x}',\Lambda_0)\equiv&\int dz_1G_{k_z}^{(A)}(z;z_1)\left[\int dz_2V_s(\mib{x}_1-\mib{x}_2)n^{(A)}(z_2) \right]u_{k_z}^0(z_1)\nonumber\\
&+\int dz_1G_{k_z}^{(A)}(z;z_1)\left[\int dz_2V(\mib{x}_1-\mib{x}_2)n^{(B)}(z_2) \right]u_{k_z}^0(z_1)\nonumber\\
&+\int dz_1G_{k_z}^{(B)}(z;z_1)\left[\int dz_2V(\mib{x}_1-\mib{x}_2)n^{(A)}(z_2) \right]u_{k_z}^0(z_1)\nonumber\\
&+\int dz_1G_{k_z}^{(C)}(z;z_1)\left[\int dz_2V(\mib{x}_1-\mib{x}_2)n^{(A)}(z_2) \right]u_{k_z}^0(z_1)\nonumber\\
&+\int dz_1G_{k_z}^{(C)}(z;z_1)\left[\int dz_2V(\mib{x}_1-\mib{x}_2)n^{(B)}(z_2) \right]u_{k_z}^0(z_1),\label{e3}
\end{align}
The 1st and 4th terms of the right hand side in eq. (\ref{e3}) are the order of $(1-s)^2$, and we neglect them. Here we consider only most divergent terms at low temperature which we used in the calculations of the perturbation theory. Then we have
\begin{equation}
\delta u_{k_z}^{(1H)}(\mib{x}',\Lambda_0)=\frac{t_0(1-|t_0|^2)}{2\pi\hbar v_F}\tilde{V}(x_1-x_2,y_1-y_2;2k_F)\ln\left(\frac{1}{s} \right)\cdot e^{ik_zz}.
\end{equation}

We put it into eq. (\ref{e1}) and perform the integrations of $x$ and $y$ directions. Then the transmitted wave within 1st Born approximation by Hartree term is written as
\begin{align}
\varphi_{\mib{k}}^{(H)}(\mib{x})=&t_0e^{ik_zz}\phi_{k_y}^0(x,y)+\varphi_{\mib{k}}^{(1H)}(\mib{x},\Lambda_0)\nonumber\\
=&t_0e^{ik_zz}\phi_{k_y}^0(x,y)\nonumber\\
&+\varphi_{\mib{k}}^{(1H)}(\mib{x},s\Lambda_0)+\alpha_1(B)t_0(1-|t_0|^2)\ln\left(\frac{1}{s} \right)\cdot e^{ik_zz}\phi_{k_y}^0(x,y)\nonumber\\
\equiv& \left\{t_0+\delta t^{(1H)}\right\}e^{ik_zz}\phi_{k_y}^0(x,y)+\varphi_{\mib{k}}^{(1H)}(\mib{x},s\Lambda_0),
\end{align}
where
\begin{equation}
\delta t^{(1H)}\equiv\alpha_1(B)t_0(1-|t_0|^2)\ln\left(\frac{1}{s} \right).
\end{equation}
Similarly, the transmitted wave within 1st Born approximation by Fock term is written as
\begin{align}
\varphi_{\mib{k}}^{(F)}(\mib{x})=&t_0e^{ik_zz}\phi_{k_y}^0(x,y)+\varphi_{\mib{k}}^{(1F)}(\mib{x},\Lambda_0)\nonumber\\
=&t_0e^{ik_zz}\phi_{k_y}^0(x,y)\nonumber\\&+\varphi_{\mib{k}}^{(1F)}(\mib{x},s\Lambda_0)-\alpha_2(B)t_0(1-|t_0|^2)\ln\left(\frac{1}{s} \right)\cdot e^{ik_zz}\phi_{k_y}^0(x,y)\nonumber\\
\equiv& \left\{t_0+\delta t^{(1F)}\right\}e^{ik_zz}\phi_{k_y}^0(x,y)+\varphi_{\mib{k}}^{(1F)}(\mib{x},s\Lambda_0).
\end{align}
with
\begin{equation}
\delta t^{(1F)}=-\alpha_2(B)t_0(1-|t_0|^2)\ln\left(\frac{1}{s} \right).
\end{equation}
Therefore the transmitted wave within the present approximation is of the form
\begin{align}
\varphi_{\mib{k}}(\mib{x})=&\left\{t_0+\delta t\right\}e^{ik_zz}\phi_{k_y}^0(x,y)+\left(\varphi_{\mib{k}}^{(1H)}(\mib{x},s\Lambda_0)+\varphi_{\mib{k}}^{(1F)}(\mib{x},s\Lambda_0) \right),
\end{align}
where
\begin{equation}
\delta t=-\alpha(B)t_0(1-|t_0|^2)\ln\left(\frac{1}{s} \right).
\end{equation}
Thus we find that $\varphi_{\mib{k}}(\mib{x})$ does not change if we reduce cutoff from $\Lambda_0$ to $s\Lambda_0$ replacing $t_0$ with $t_0+\delta t$ at the same time. The transmission probability changes:
\begin{equation}
{\cal T}_0=|t_0|^2 \to {\cal T}_0+\delta{\cal T}\equiv|t_0+\delta t|^2,
\end{equation}
and we obtain eq. (\ref{rgprob1}), where in stead of $\Lambda_0$, we use the energy $E_0\equiv\hbar v_F\Lambda_0, E\equiv\hbar v_Fs\Lambda_0$.
\end{appendix}

\end{document}